%% file: main-IEEE.tex
\documentclass[conference]{IEEEtran}
\IEEEoverridecommandlockouts
\usepackage{cite}
\usepackage{amsmath,amssymb,amsfonts}
\usepackage{algorithmic}
\usepackage{graphicx}
\usepackage{textcomp}
\usepackage{xspace}
\usepackage{comment}
\usepackage{booktabs}
\usepackage{enumitem}
\usepackage{xcolor}
\usepackage{url}
\usepackage{listings}
\usepackage{subfig}
\usepackage[hidelinks]{hyperref}
\def\BibTeX{{\rm B\kern-.05em{\sc i\kern-.025em b}\kern-.08em
    T\kern-.1667em\lower.7ex\hbox{E}\kern-.125emX}}

\newcommand{\tool}{SODA\xspace}

\makeatletter
\newcommand{\newlineauthors}{%
  \end{@IEEEauthorhalign}\hfill\mbox{}\par
  \mbox{}\hfill\begin{@IEEEauthorhalign}
}
\makeatother
    
\begin{document}

\title{Disaggregated Memory with SmartNIC Offloading: a Case Study on Graph Processing}

\author{\IEEEauthorblockN{Jacob Wahlgren}
\IEEEauthorblockA{\textit{KTH Royal Institute of Technology}\\
Stockholm, Sweden \\
jacobwah@kth.se}
\and
\IEEEauthorblockN{Gabin Schieffer}
\IEEEauthorblockA{\textit{KTH Royal Institute of Technology}\\
Stockholm, Sweden \\
gabins@kth.se}
\and
\IEEEauthorblockN{Maya Gokhale}
\IEEEauthorblockA{\textit{Lawrence Livermore National Laboratory}\\
Livermore, USA \\
gokhale2@llnl.gov}
\newlineauthors
\IEEEauthorblockN{Roger Pearce}
\IEEEauthorblockA{\textit{Lawrence Livermore National Laboratory}\\
Livermore, USA \\
pearce7@llnl.gov}
\and
\IEEEauthorblockN{Ivy Peng}
\IEEEauthorblockA{\textit{KTH Royal Institute of Technology}\\
Stockholm, Sweden \\
ivybopeng@kth.se}
}

\maketitle

\begin{abstract}
Disaggregated memory breaks the boundary of monolithic servers to enable memory provisioning on demand. Using network-attached memory to provide memory expansion for memory-intensive applications on compute nodes can improve the overall memory utilization on a cluster and reduce the total cost of ownership. However, current software solutions for leveraging network-attached memory must consume resources on the compute node for memory management tasks. Emerging off-path smartNICs provide general-purpose programmability at low-cost low-power cores. This work provides a general architecture design that enables network-attached memory and offloading tasks onto off-path programmable SmartNIC. We provide a prototype implementation called \tool on Nvidia BlueField DPU. \tool adapts communication paths and data transfer alternatives, pipelines data movement stages, and enables customizable data caching and prefetching optimizations. We evaluate \tool in five representative graph applications on real-world graphs. Our results show that \tool can achieve up to 7.9x speedup compared to node-local SSD and reduce network traffic by 42\% compared to disaggregated memory without SmartNIC offloading at similar or better performance.
\end{abstract}

\begin{IEEEkeywords}
SmartNIC, Disaggregated Memory, Fabric-Attached Memory
\end{IEEEkeywords}

\input{intro}

\input{background}

\input{design}

\input{impl}
\input{setup}

\input{results}

\input{related}

\input{conclusion}

\section*{Acknowledgment}
This work was performed in part under the auspices of the U.S. Department of Energy by Lawrence Livermore National Laboratory under Contract DE-AC52-07NA27344 (LLNL-CONF-869710). Funding from LLNL LDRD project 24-ERD-047 was used in this work. This research is also supported by the Swedish Research Council (no. 2022.03062).

\bibliographystyle{IEEEtran}
\bibliography{main}

\end{document}

%% file: intro.tex
\section{Introduction}
Large-scale computing clusters need to facilitate a diverse mixture of workloads with varying resource demands. More memory- and data-intensive workloads, such as deep learning and graph processing applications~\cite{shun2013ligra,zhang2018graphit}, are using clusters to fulfill their high computing needs. However, recent studies show that clusters with large node-level memory resources can result in significant resource under-utilization~\cite{zivanovic2017main,panwar2019quantifying,peng2020memory,michelogiannakis2022case,li2023analyzing}. Moreover, memory is becoming an increasingly important component of the total ownership cost and thus equipping compute nodes with large DRAM may become prohibitively expensive. Systems like pre-exascale CORAL systems~\cite{vazhkudai2018design} use node-local NVMe SSD to augment DRAM capacity and enable memory- and data-intensive applications in a cost-effective way. However, such solutions are dependent on the underlying infrastructure. Alternative solutions~\cite{panwar2019quantifying,peng2020memory} leverage network-attached nodes to provision memory resources when compute nodes need to support memory-intensive applications, resulting in disaggregated memory and compute.

\begin{figure}
    \centering
    \includegraphics[width=\linewidth]{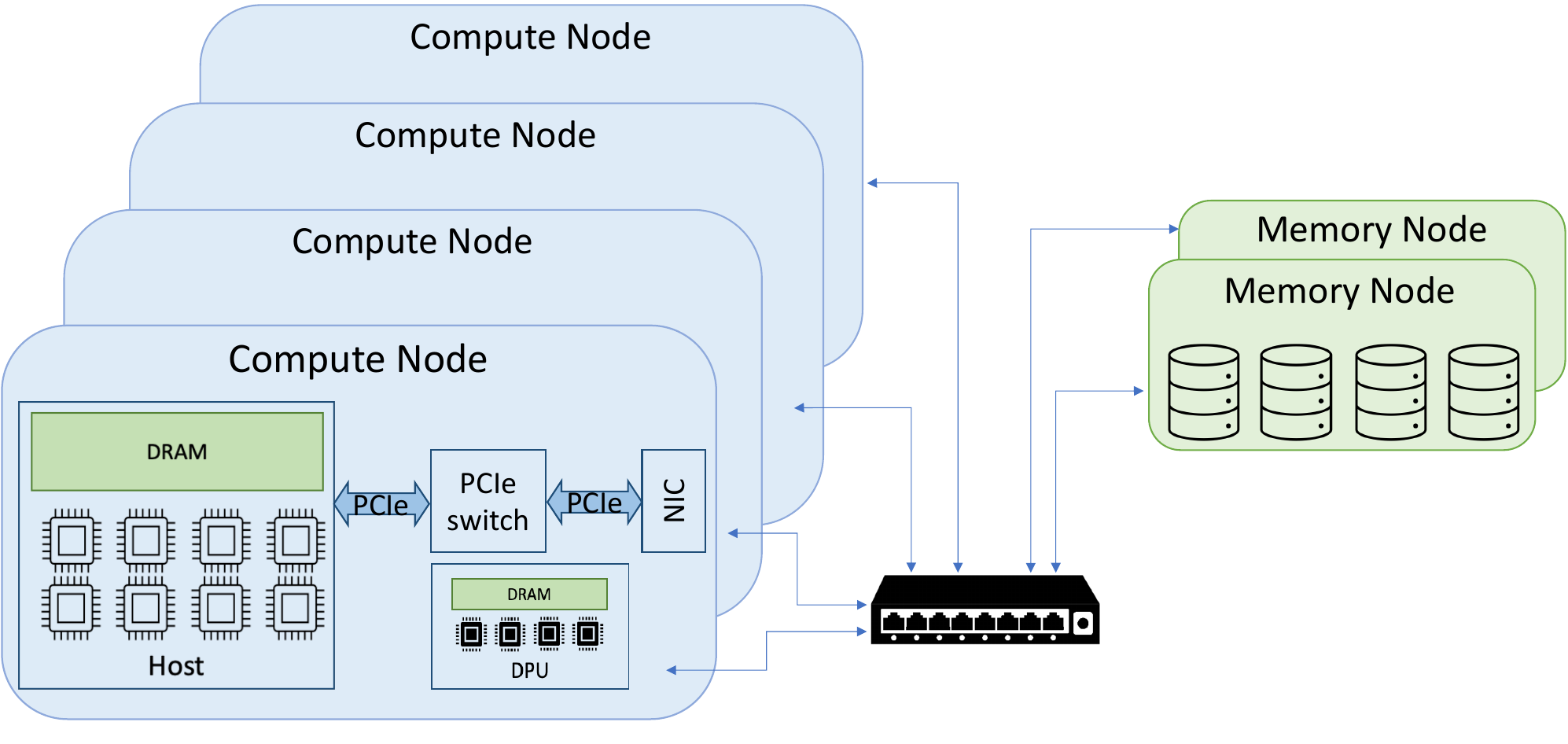}
    \caption{A cluster composed of compute nodes and memory nodes. Each compute node consists of the host CPU and an off-path SmartNIC with DPU. Each memory node is equipped with massive memory resources.}
    \label{fig:arch}
\end{figure}

The recent development in fast interconnects, including cache-coherent Compute Express Link (CXL) protocol and RDMA-enabled networking, enables rack-scale memory disaggregation~\cite{lim2009disaggregated,gu2017efficient,li2023pond,wahlgren2023quantitative} as a cost-effective and scalable solution by provisioning memory resources on-demand. In a disaggregated design, as exemplified in Figure~\ref{fig:arch}, the compute nodes are equipped with moderate memory capacity, which can be further augmented by network-attached memory nodes when needed. Thus, the ownership cost can be kept low compared to systems composed of large-memory compute nodes. Since multiple compute nodes can share memory nodes, the overall resource utilization is also improved. Additionally, compute nodes and memory nodes may be upgraded independently. Existing software solutions~\cite{lim2009disaggregated,gu2017efficient,zahka2022fam} for enabling disaggregated memory, either via OS-based solutions or domain-specific programming approaches, need specific memory management tasks. These tasks run on compute nodes to manage and optimize data movement between compute nodes and network-attached memory.

In this work, we explore off-path general programmable SmartNIC to offload these memory management tasks for enabling disaggregated memory. We propose a runtime library called \tool (\textbf{S}martNIC-\textbf{O}ffloaded \textbf{D}is\textbf{A}ggregated memory). \tool is implemented atop Nvidia's BlueField DPU, for coordinating and pipelining data movement between compute nodes and memory nodes. Off-path SmartNIC~\cite{wei2023characterizing}, as illustrated in DPU in Figure~\ref{fig:arch}, refers to a separate set of cores and memory packed into a system-on-chip (SoC) attached to the network interface card (NIC) via a PCIe switch. Thus, the SoC can send and receive packets independently from the main host CPU, without intervening the OS. Therefore, some host-side tasks for managing data movement between the compute and memory nodes, may be offloaded onto the off-path SoC. As compute and memory on SmartNIC SoC are more power efficient and cost-effective than high-end host processors, offloading tasks onto SmartNIC SoC can save resources on the host for resource-demanding computational tasks.

We design \tool to enable memory-intensive applications on memory-limited nodes by augmenting memory from fabric-attached memory (FAM). \tool provides simple allocation APIs for selecting and transforming memory objects in an existing application into FAM-backed memory objects with minimal modifications. Memory accesses to these FAM-backed memory objects are transparently translated into network requests and sent to the memory node. As a runtime solution, the application can have explicit control over FAM-backed memory objects, unlike OS-level solutions. For instance, it supports applications to specify the amount of memory resources to be backed from the host and FAM, respectively. Moreover, \tool supports transparently offloading memory management tasks onto off-path SmartNICs when available.

At a high level, \tool consists of three components that reside on the host, the DPU (attached to off-path SmartNIC), and the memory node, respectively. Internally, the DPU component coordinates, merges, and pipelines data transfers between the compute node and memory node. Multiple independent processes on one compute node can share the \tool service on DPU to improve resource utilization. Furthermore, \tool supports two data caching schemes on the DPU to adapt to application characteristics to reduce data movement over the network. We evaluated \tool in five graph applications on real-world graph datasets. \tool achieves up to 7.9x speedup compared to node-local NVMe and up to 42\% reduction in network traffic at similar or faster performance compared to non-offloading solutions. We summarize our contributions in this work as follows.

\begin{itemize}[leftmargin=*,topsep=5pt]
    \item We describe the \tool design for enabling network-attached memory and offloading tasks onto off-path SmartNIC.
    \item We leverage characterization and analytical modeling to guide NUMA-awareness and caching optimizations.
    \item We provide a prototype implementation on the Nvidia BlueField DPU and RDMA Verbs API.
    \item We evaluate \tool in five graph applications on real-world graphs in single-and multi-process scenarios.
    \item \tool achieves up to 7.9x speedup compared to node-local SSD and up to 42\% reduction in network traffic.
\end{itemize}

%% file: background.tex
\section{Background and Motivation}

\subsection{Disaggregated Memory}
Disaggregated memory is a system design that separates memory resources from compute resources. In contrast to monolithic servers where compute and memory are tightly coupled within a node's boundary, memory-compute disaggregation could improve resource utilization and system flexibility. In today's HPC systems, the resources in a compute node are typically assigned to a single job, regardless of its actual utilization. As jobs have diverse needs on their compute and memory resources, memory resources on a cluster are often significantly underutilized~\cite{panwar2019quantifying,peng2020memory}. Disaggregated memory can support memory resources to be provisioned on-demand~\cite{lim2009disaggregated}, reducing memory utilization.

Memory disaggregation is mostly implemented via fast interconnect technologies, such as cache-coherent CXL~\cite{li2023pond,wahlgren2022evaluating} and fast networks with RDMA support~\cite{lim2009disaggregated,gu2017efficient}. Regardless of the specific interconnects in use, a common \textit{pooled} design, as illustrated in Figure~\ref{fig:arch}, consists of compute nodes and memory nodes (pools). The compute nodes are high-end computing units, e.g., multi-core CPU and GPU, and moderate DRAM capacity that can support common workloads. When compute nodes need to facilitate memory-intensive workloads, they can expand the memory capacity by accessing FAM pools, which are equipped with massive DRAM capacity. As a memory node can serve multiple compute nodes and memory is provisioned on-demand, such system design can improve resource utilization~\cite{lim2009disaggregated}. 

Existing solutions for using network-attached memory as disaggregated memory~\cite{lim2009disaggregated,gu2017efficient,zahka2022fam} need software frameworks for handling memory management, consistency, requesting and transferring data over the fabric, resilience, etc. At the bare minimum, a framework for utilizing disaggregated memory must track access to remote memory, transfer data from remote memory, write back dirty data to remote memory, handling consistency. Depending on whether in application-level or OS space and the target workload and system characteristics, different optimizations may be enabled in different solutions. All these memory management tasks need to use resources when running on a compute node, consequently reducing the high-end compute resource available to the application itself. Recent off-path programmable smartNIC comes with low-cost computing resources. As they have DPU attached to the PCIe switch in a compute node (as shown in Figure~\ref{fig:arch}), they can potentially offload some tasks from the host to its DPU and free up resources on the host.

\subsection{SmartNIC Technologies}
SmartNICs refer to advanced network interface cards (NICs) that have the computing power to offload some tasks from the host. Depending on their programmability, SmartNIC can be categorized as either fixed or general programmable. A SmartNIC may be only able to offload specific network functions like packet processing. Or, when equipped with programmable processors like Field Programmable Gate Arrays (FPGAs) or System on Chips (SoCs), programmers can offload general computing tasks onto SmartNIC. General programmable SmartNICs are becoming more accessible, as represented by Intel's FPGA-based SmartNIC and Nvidia's ARM-based BlueField SmartNIC. In this work, we explore offloading tasks for managing disaggregated memory onto SmartNIC, and thus general programmable SmartNICs are considered. 

General programmable SmartNICs have cores and memory in SoCs attached to the NIC. Within a compute node, the host, SoC, and NIC are connected by a PCIe switch, as shown in the inset in Figure~\ref{fig:arch}. Network packets from the host can bypass the SoCs, i.e., off-path SmartNICs. In contrast, fixed-function SmartNICs are often en route for all network packets. Due to the power and thermal constraints, SmartNICs use low-power cores. Thus, they are ideal for tailored tasks that require low computing power. For off-path SmartNICs, when getting data from the host or network, due to the PCIe switch, at least two hops on PCIes are required. However, compared to network-attached nodes, utilizing resources on SoC within a node provides better isolation and reduces exposure to network noises on the public cloud and data centers~\cite{de2022noise}.  

\subsection{Nvidia BlueField SmartNIC}
In this work, we target off-path smartNIC represented by the Nvidia BlueField SmartNIC DPU, where the programmable compute resources are separated from the NIC's packet processing resources~\cite{liu2021performance,karamati2022smarter,wei2023characterizing}. The DPU contains multiple low-power ARM cores and DRAM. Both host and DPU run a stand-alone Linux operating system and thus they can be considered as two separate endpoints. Because of this `separate-host' mode, the programmability of the off-path SmartNICs is improved compared to low-level on-path SmartNICs, and existing codes can be simply compiled and run on the DPU without porting efforts. Both the host and DPU can issue RDMA operations to the NIC using regular programming interfaces such as \texttt{ibverbs}. To use specialized features on the SmartNIC, such as the compression engine and DMA controller, programmers need to use APIs provided in Nvidia's DOCA SDK. There are three generations of BlueField DPUs, with major differences in the number of ARM cores, DRAM on DPU, and PCIe generation. Though their peak capacity may differ, their architecture and position in the disaggregated system remain unchanged and thus, the design proposed in this work uses general programming off-path SmartNIC. However, as SmartNICs are in active development, solutions atop SmartNICs should be flexible for adapting to new generations.

%% file: design.tex
\section{\tool Design}
\label{sec:design}
We propose \tool as a runtime library for memory-intensive applications to leverage network-attached memory. \tool supports transparently offloading tasks of managing data movement between compute nodes and memory nodes onto off-path general programmable SmartNICs.

\begin{figure}
    \centering
    \includegraphics[width=\linewidth]{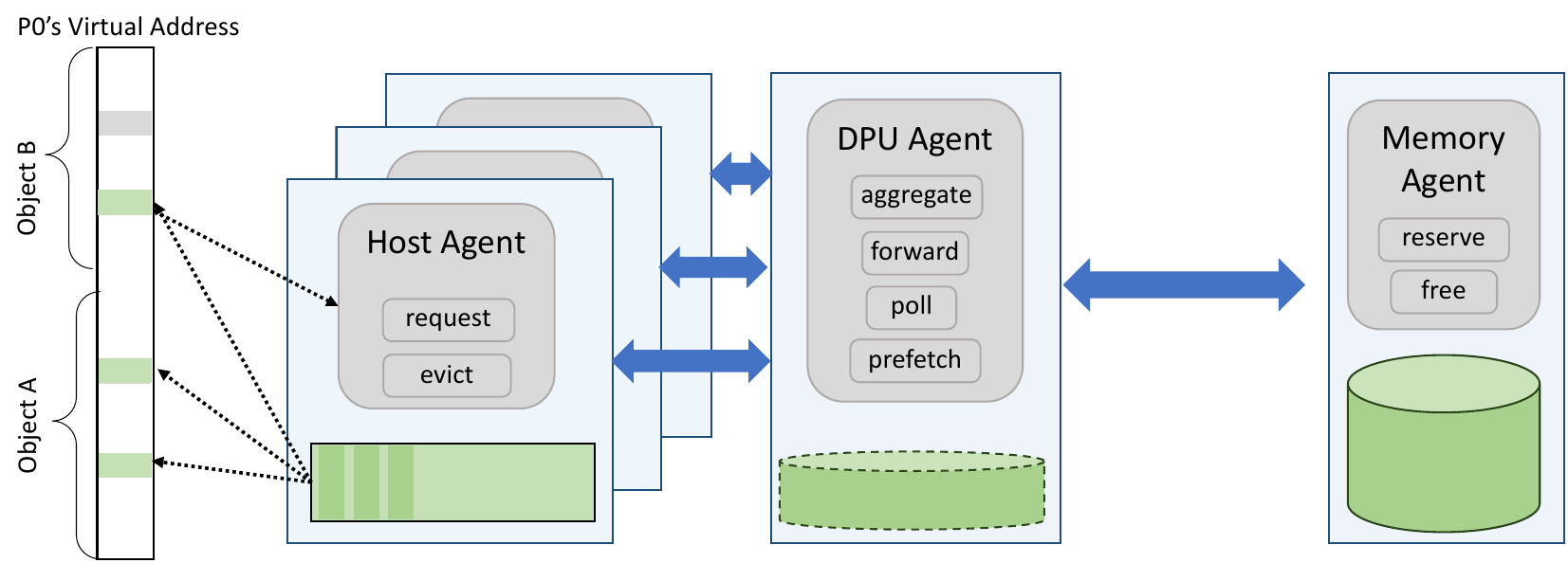}
    \caption{\tool consists of agents on the host, SmartNIC SoC (DPU), and memory server. Memory objects in the application process's virtual space can be backed by network-attached memory. \tool agents transparently handle tasks for memory management and data movement for the application.}
    \label{fig:design}
\end{figure}

\tool consists of three agents, running on the host, DPU, and memory node, respectively. Figure~\ref{fig:design} depicts their main components and interactions. The \textbf{host agent} manages a memory buffer for staging data from the memory node. Its main tasks include issuing requests for data and evicting data when the buffer is full. The \textbf{DPU agent} is tasked with receiving and processing requests from the host, aggregating and forwarding requests to the memory node, managing and optimizing data movement between the compute and memory nodes. It also leverages its local DRAM to prefetch and cache data. A DPU agent may handle multiple host agents on a compute node. The \textbf{memory agent} is deployed on the memory node and it only handles simple tasks like reserving and freeing memory resources.

Although OS-based system software solutions, such as paging and swapping, can utilize network-attached memory transparently, they have a dependency on kernel versions, require privileged access on a cluster, and have system-wide impacts. Users of data centers and HPC systems typically have no privilege in configuring the OS kernel. Thus, we choose a user-space design.

To reduce porting efforts, we design \tool to interface with applications only through memory objects. In particular, an application can use \tool APIs to allocate FAM-backed data objects. Internally, \tool monitors memory accesses to these FAM-backed objects and leverages its agents to coordinate data transfer from network-attached memory transparently for the application. A FAM-backed object is a contiguous memory region in a process's virtual address space, just like usual memory allocations, as illustrated in memory objects \textit{A} and \textit{B} in Figure~\ref{fig:design}. During the allocation, the memory agent reserves sufficient memory resources, which are then mapped by the host agent into the application's virtual address space. The host agent maintains the metadata and mapping between FAM-backed objects and memory nodes.

The host agent monitors access to FAM-backed memory regions. It also manages a memory buffer shared by all FAM-backed objects for caching data in the host's DRAM. When the application accesses a FAM-backed object, if the accessed data is not resident in the buffer, the host agent issues requests for fetching data from network-attached memory. As the memory buffer is smaller than FAM-backed objects, when the buffer is full, the host agent will evict dirty data back to network-attached memory to free up the buffer. Our design chooses to use a unified buffer for all FAM-backed objects and employ an LRU policy to manage the buffer to ensure the local buffer is distributed to FAM-backed objects as needed. \tool supports the size of the buffer to be controlled at the application level as different applications may exhibit different sensitivity to the buffer size.

The host agent manages the buffer in equal-sized data chunks. Each data chunk is the minimum unit of data movement between the compute nodes and memory nodes. \tool is designed to support highly concurrent data accesses in multi-threaded parallel applications. Read and write accesses are aggregated by data chunks to reduce the overhead. \tool supports the size of data chunks to be controlled at the application level.

Finally, we employ NUMA-awareness when placing the host agent on the compute node to optimize performance. Today, hosts commonly have two or more NUMA nodes. As NIC is connected to one NUMA node, data transfer to and from this NUMA node would exhibit higher performance than from other NUMA nodes. Therefore, when allocating the communication buffer for receiving data, \tool binds the communication buffer to the NUMA node closest to NIC to ensure that data transfer between compute nodes has optimal performance in bandwidth and latency and also reduces performance variability, compared to the default behavior.

The DPU agent receives two types of requests from the host agent. First, when the application accesses a data chunk that is not resident in the memory buffer, the host agent sends a request with the corresponding metadata information to the DPU agent. Second, when the memory buffer is full and even an evicted data chunk is dirty, the host agent will evict the dirty data chunk. In this write-back process, the host agent sends the data to the DPU agent and returns immediately. Without offloading to DPU, the eviction process is synchronous until all data reaches the memory node. To avoid eviction on the critical path, we employ a proactive eviction policy that is triggered when the buffer reaches a threshold load factor.

Memory coherence becomes a challenge in a scenario with multiple clients mapping the same writable FAM object. Coherency solutions include snoop protocols based on a shared transaction bus and directory protocols with a central metadata service~\cite{ewais2023disaggregated}. However, due to the limited scalability and the complexity of coherency protocols, we restrict \tool writable mappings to single clients only.

The DPU agent maintains the metadata of FAM data objects, memory nodes, and their mapping. When host requests arrive, it checks the metadata to compose corresponding operations from the network-attached memory. It then forwards the requests to the memory node and actively polls for the completion of the requests. Once the data is fetched, the DPU agent will move it to the memory buffer on the host component. In this stage, the design employs a zero-copy data transfer strategy by using the same buffer on DPU for receiving data from the memory node and moving to the host side without data copy. These various tasks for managing metadata, data movement, and, coordination, could execute on the host side, as in previous works, but they will consume the compute resources useful for running applications. The \tool design supports offloading these tasks onto programmable SmartNICs.

The DPU agent is able to handle multiple processes on a compute node since it does not require any process-specific management. This DPU sharing is fully transparent from the client's perspective. However multiple co-located processes further multiply the number of requests and increase the opportunity for coordination in the DPU agent.

As multi-threaded applications generate highly concurrent requests, we further propose two novel optimizations to harness multiple cores on the DPU agent to improve performance.

\textbf{Task Aggregation.} The DPU agent receives many concurrent requests from the host agent in multi-threaded applications. Processing each request sequentially leads to stalls since available hardware parallelism is not utilized. Also, each request incurs an overhead in sending commands to the NIC. To overcome these inefficiencies, the DPU agent aggregates concurrent requests into a \textit{task batch}. All network operations in one batch are processed in parallel. This batching optimization avoids queuing delays and reduces the NIC overhead~\cite{kalia2016design}. The aggregation task needs to use memory for maintaining the state of all requests in a task batch. However, as the metadata is less than 1~kb per request, this overhead is negligible on current generations of Nvidia BlueField.

Task aggregation is beneficial when there is a high concurrency of requests. However, aggregating requests incurs one extra step in each request, thus increasing the latency of a single request. Thus, this aggregation optimization should only be used for highly concurrent parallel applications, which are common in HPC systems. These multi-threaded applications typically leverage multiple hardware cores and threads to issue highly concurrent requests. The aggregation optimization are particularly useful in the write-back process, where the DPU agent can combine multiple requests before writing back to the corresponding FAM regions. 

\textbf{Asynchronous Request Forwarding.} We leverage multiple cores on the DPU to split the request receiving and forwarding into a pipeline. When the DPU agent forwards a request to the memory node, the DPU agent needs to wait for its completion. This blocking operation limits its scalability when more new tasks are waiting. To increase the throughput, request forwarding is pipelined in two separate threads by asynchronously handling the communication to the memory node. One thread is responsible for interacting with the host agent in receiving requests, looking up their metadata, composing specific operations to the memory node, and initiating server operations. The other thread is dedicated to polling for responses from the memory node operations and then staging the data to the host agent's memory buffer.

\input{dpu-cache}

%% file: dpu-cache.tex
\subsection{Data Caching in DPU}
\label{sec:cache}
Data movement over the network is a key challenge of disaggregated memory. Network congestion and noise may cause performance variability, and large network traffic can lead to contention with other workloads and limit the available network throughput~\cite{de2022noise}. To address the data movement challenge, we propose using the SmartNIC for caching remote memory locally within a compute node. If a request hits the data cached in the DPU agent, it can return to the host agent immediately, thus avoiding at least two hops over the network. If it misses, the request follows the same path as before to the memory node. The main trade-off between the gain of network traffic reduction and the overhead of extra checks of the cache depends on the hit rate to the cache in DPU and the memory system on the DPU. As the latency to look up data in DPU's DRAM is typically in hundreds of nanoscales while latency over the network is in several microseconds, a request that can be fulfilled by DPU cache can finish in shorter latency than going through the network. When the number of concurrent requests is high, another critical factor is the bandwidth between the host and DPU, as compared to the network bandwidth. Depending on the generation of SmartNICs, the bandwidth between the host and DPU can be limited by different generations of PCIe. Also, a platform may use low-bandwidth commodity or high-performance networks, influencing the peak network bandwidth. Finally, depending on the workload of a system, the network bandwidth is shared among co-running jobs. Therefore, we design the DPU caching as an optional module that can be enabled based on system characteristics. When the hit rate to the DPU cache is high, caching on DPU will always reduce network traffic between compute node and memory node.

We exploit the programmability of off-path SmartNIC to explore two customizable caching strategies, static and dynamic caches. We design the caching on the DPU to be adaptive to the monitored hit rate. A caching strategy with low accuracy and hit rate may lead to increased traffic to the memory node. Caching on DPU can be disabled when it is not beneficial to the workload. Moreover, applications can develop application-specific caching policies to be deployed onto the DPU agent.

\textbf{Static Caching} leverages application-specific knowledge to place selected data chunks into the DPU cache. Static caching requires low overhead as the cached data is not updated. By extending the metadata on the host agent, \tool can determine whether a page is cached in DPU or choose to bypass it. Therefore, the static caching strategy can achieve a 100\% hit rate on the DPU cache. However, it has limitations from the DPU's memory subsystem. The DPU memory capacity is typically small, and depending on the generation of SmartNIC, DRAM on DPU may be slow. Thus, this caching strategy relies on the ability to identify small memory regions with very high access density.

\textbf{Dynamic Caching} monitors data access patterns at runtime to prefetch data that is likely to be used in the future. Based on accesses to the DPU cache, the prefetcher loads adjacent data chunks from the memory node and stages them on the DPU cache, which occurs off the critical path. Moreover, the larger transfer size avoids the overhead of several smaller transfers~\cite{kalia2016design}. The performance of dynamic caching relies on the accuracy of the prefetching scheme. For workloads with high hit rates, most latency-critical on-demand fetches can be served locally from the DPU cache, i.e., effectively converting a majority of on-demand data transfers into background transfers off the critical path. Compared to static caching, dynamic caching is adaptable to runtime behaviors, such as changed hot regions throughout the execution, and flexible capacity in the DPU. However, its adaptivity comes at the cost of high maintenance overhead. 

In general, we use a larger cache entry size than page size to effectively realize the prefetching mechanism. The ratio between page and cache entry size is a trade-off between hit rate, accuracy, and read amplification. The optimal value will depend on the access pattern of the workload, which is why we leave these values as tunable parameters.

We derive an analytical model to guide the selection of caching strategies in \tool on a target platform. In the baseline case, data is fetched directly from the memory node, the time $T$ to fetch a data chunk of $s$ bytes with a network bandwidth of $B_{net}$ is
\begin{equation}\label{eq:T_0}
    T = \frac{s}{B_{net}}.
\end{equation}
The time to fetch the same data chunk using dynamic caching ($T_d$) depends on the bandwidth between the host and the DPU $B_{intra}$ as well as the hit-rate $h$. %
\begin{equation}
    E[T_d] = \frac{s}{B_{intra}} + (1-h)\frac{s}{B_{net}}
\end{equation}
Dynamic caching is beneficial if 
\begin{equation}\label{eq:dyn_benefit}
    E\Big[\frac{T}{T_d}\Big]>1 \iff h > \frac{B_{net}}{B_{intra}}.
\end{equation}
Given the network-to-intra bandwidth ratio ($R=\frac{B_{net}}{B_{intra}}$) on a target platform, we are able to deduce the required hit rate.  For a $R$ of 1:2, we need a hit rate above 50\% and for a $R$ of 1:3, we only need a hit rate above 33\%. Note that this model only considers the idle peak network bandwidth, while the actual network bandwidth depends on the system load at runtime. %

%% file: impl.tex
\section{Implementation on the Testbed}
\label{sec:impl}
In this section, we introduce the implementation of communication strategy and protocols, caching, and APIs of \tool. \tool is implemented in C++, using ibverbs RDMA\footnote{\url{https://github.com/KTH-ScaLab/SODA}}.

The implementation is based on benchmarks of our testbed. It consists of nodes with dual-socket AMD EPYC 7401 processors with a total of 48 cores running at 2.0~GHz and 256~GB DDR4 memory of 16 channels running at 2400 MT/s. Each node is also equipped with an Nvidia BlueField-2 DPU. The DPU has an ARM Cortex A-72 processor with 8 cores and 16~GB DDR4 memory of one channel running at 3200~MT/s. The DPU is configured to run in separated host mode so that network packets bypass the off-path DPU SoC. The network between compute nodes and memory nodes is using RoCE on 100~Gb/s Ethernet. 

\subsection{Selecting Communication Strategy}
\label{sec:char}
Between host and DPU, we compare two data transfer options: RDMA through ibverbs and DMA through the DOCA SDK. RDMA supports both one-sided operations, i.e., write and read, and the two-sided send operation. DMA supports read and write issued from the DPU agent. We benchmark the performance of RDMA using linux-rdma's perftest\footnote{https://github.com/linux-rdma/perftest}. We developed a custom benchmark code to measure the DMA performance using Nvidia DOCA SDK (version 2.1.0). Although these results are obtained on a specific platform, our benchmarking approach is applicable to different systems for adapting implementation choices.

\begin{figure}
    \centering
    \includegraphics[width=\linewidth]{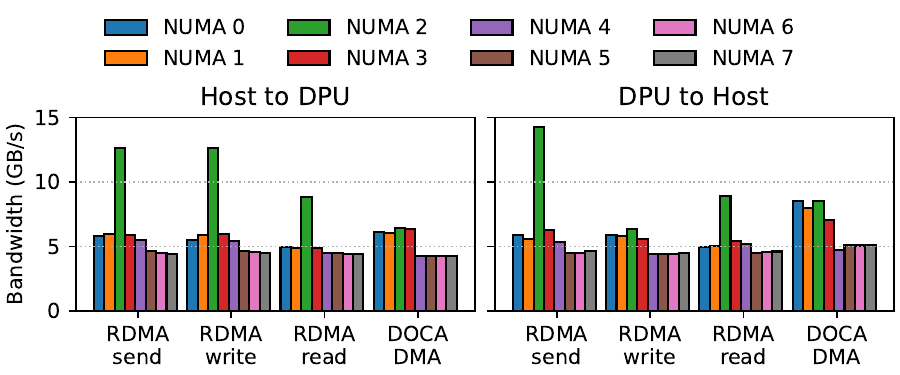}
    \caption{Performance variation when using different NUMA nodes in the host memory at message size 64~KB.}
    \label{fig:numa_bars}
\end{figure}

Figure~\ref{fig:numa_bars} shows a strong NUMA effect on intra-node communication between the host and DPU. As expected, the best performance is offered on NUMA node~2, where the NIC is attached. The performance difference to other NUMA nodes varies significantly, indicating that NUMA-aware host agent placement is critical. We implement the NUMA-aware placement in \tool using Linux's \textit{libnuma} library.

\begin{figure}
    \centering
    \includegraphics[width=\linewidth]{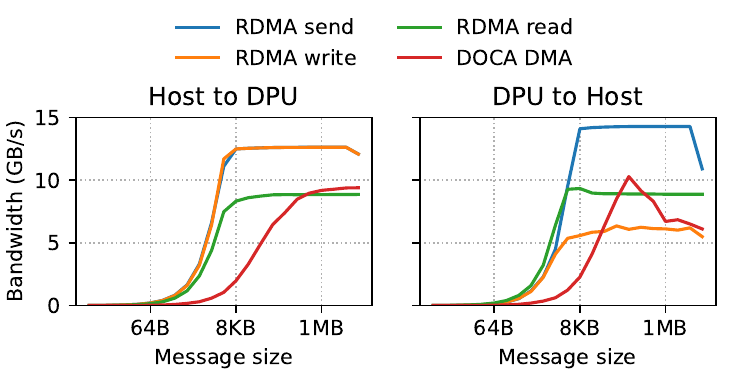}
    \caption{Performance of different intra-node communication options (using the fastest NUMA configuration).}
    \label{fig:intra_comm}
\end{figure}

The communication performance is also impacted by the size of data chunks on the host agent. Figure~\ref{fig:intra_comm} shows the bandwidth as a function of size, for RDMA and DMA, respectively. For RDMA, the bandwidth reaches a plateau at 4--8~KB message size. The peak bandwidth varies by RDMA operation and direction. The fastest is DPU to host SEND, with 14.3~GB/s, followed by host to DPU SEND and WRITE with 12.6~GB/s. READ peaks around 9~GB/s, and the slowest is DPU to host WRITE with 6~GB/s. Like with RDMA, DMA bandwidth also depends on the target host NUMA node. It has up to 10.3~GB/s write at 64~KB and 9.4~GB/s read at 8~MB. The read bandwidth increases with message size, with 7.4~GB/s at 64~KB, and flattening out around 512~KB with 9.0~GB/s. The write bandwidth peaks at 64~KB and then decreases again down to 6.1~GB/s at 8~MB. As RDMA yields the same or better performance compared to DMA in most cases. RDMA also has greater flexibility as it can be issued from both the host and DPU sides. Also, DMA requires a separate control path, as the host cannot detect when DMA has been completed on the DPU. Therefore, we implement \tool communication using RDMA and configure the data chunk to 64~KB.


\subsection{RDMA Communication Protocol}
\tool uses an RPC-based control plane protocol to manage setup and teardown of RDMA queue pairs (QPs), loading region data, etc. The host agent maintains multiple QP for communication with the DPU agent and memory node. Using multiple independent QPs avoids locking and improves NIC parallelism compared to using a single shared QP~\cite{kalia2016design}. The data plane is implemented in two RDMA-based protocols, i.e., one-sided and two-sided protocols. Table~\ref{tab:reqs} summarizes their request formats. %

The one-sided protocol utilizes the one-sided RDMA primitives to read data directly from the remote memory, where the remote endpoint is passive. Thus, it requires the full region data to be already present in the remote endpoint's memory when the request is issued. The one-sided protocol is used to access server data and in the static cache strategy. However, for dynamic caching, the one-sided protocol cannot be used because the DPU must actively do a cache lookup step. %

\begin{table}
    \centering
    \subfloat[Read request.]{
    \begin{tabular}{ll}
    \toprule
         Field & Bits \\
         \midrule
         \texttt{region\_id} & 16 \\
         \texttt{page\_offset} & 48 \\
         \texttt{dest\_addr} & 64 \\
         \texttt{size} & 32 \\
         \texttt{dest\_rkey} & 32 \\
         \bottomrule
    \end{tabular}
    }~~~
    \subfloat[Write request.]{
    \begin{tabular}{ll}
    \toprule
         Field & Bits \\
         \midrule
         \texttt{region\_id} & 16 \\
         \texttt{page\_offset} & 48 \\
         \texttt{size} & 32 \\
         \texttt{data} & variable \\
         \bottomrule
         \\         
    \end{tabular}
    }
    \caption{Request format in \tool the two-sided protocol.}
    \label{tab:reqs}
\end{table}

The two-sided protocol uses RDMA send primitives and is used when the DPU must do in-line processing of requests, such as in dynamic caching. Immediate data is used to specify the request type, either read or write. The response to a read request can use either a send operation or a one-sided write operation. In the case of write, the \verb|dest_addr| and \verb|dest_rkey| in the request are used for writing the response. On our testbed, the send operation is selected (see Figure~\ref{fig:intra_comm}).

The DPU uses a shared RDMA receive queue for receiving multiple incoming requests into a communication buffer, enabling multiplexing of several requesting endpoints into the same communication buffer. With task aggregation, multiple forwarding requests are sent as a group using doorbell batching to reduce NIC overhead~\cite{kalia2016design}. %

\subsection{Implementing Caching Strategies}
The DPU agent maintains two data structures: \textit{Recent List} and \textit{Cache Table} for enabling dynamic caching. Both data structures support concurrent accesses from multiple threads on the DPU.

The recent list maintains a history of recent accesses used for prefetching. It is implemented in a ring buffer storing the ids of the 128 most recently requested pages. For each new request, the DPU agent pushes the requested id to the head of the list. The tail element is overwritten if the list is full. Since the recent list is accessed by threads for processing requests and threads for prefetching, it needs to be thread-safe. As the critical section is small, contention is unlikely, and we use a mutex for simplicity. A condition variable is used to enable prefetching workers to wait for new requests. Alternatively, a lock-free data structure using atomics could be used to allow scaling to a higher number of DPU threads.

The cached table is used to cache data. The cache table data is stored in a fixed-size registered memory region, which enables zero-copy request fulfillment from the cache. To enable efficient lookups, the cache table uses a hash table to map from requested ids to cache entries. To minimize overhead, a random cache eviction policy is used. A simple mutex could be used to protect the cache table from concurrent access and modification. However, it would limit the cache throughput. Instead, we use a refcount on each cache entry to track the number of outstanding request fulfillment on this slot. An entry with a positive refcount is prevented from being evicted. A mutex is still used to protect the cache table metadata, but it is not required to be held during the full request processing.

\begin{figure}[bt]
    \centering
    \includegraphics[width=\linewidth]{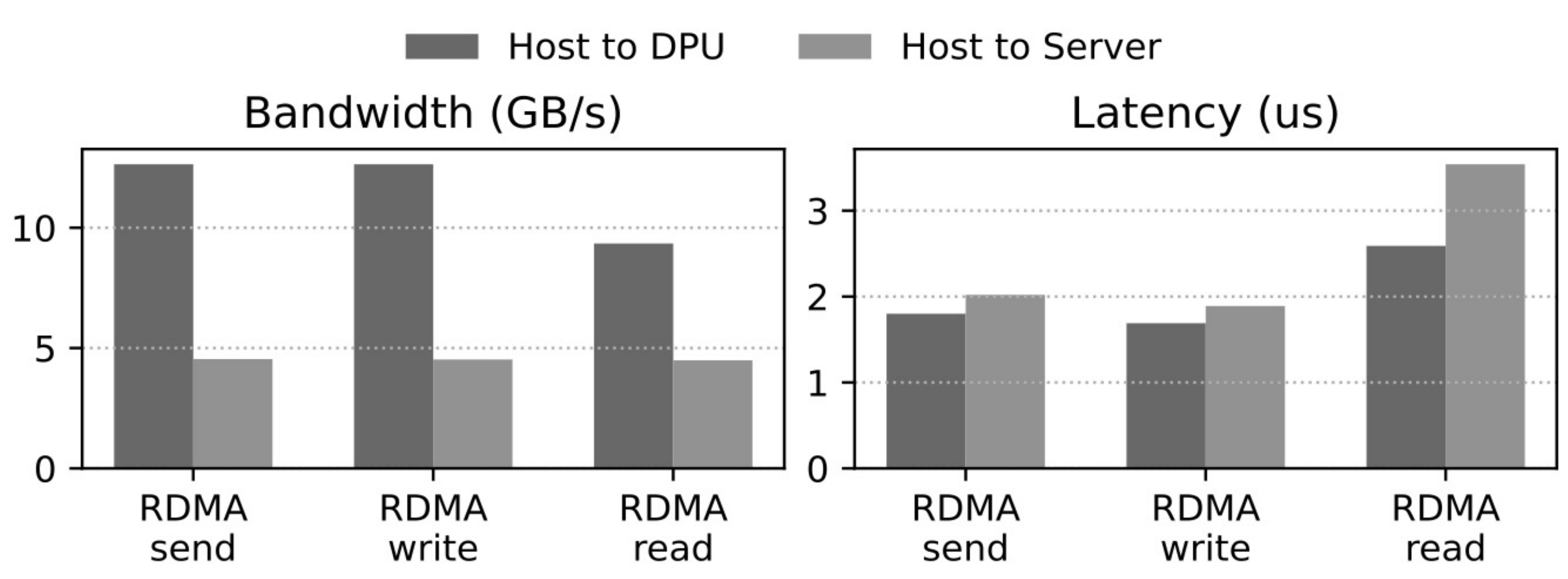}
    \caption{Comparison of performance between intra-node and inter-node communication on the testbed.}
    \label{fig:network_compare}
\end{figure}

As introduced in Section~\ref{sec:design}, we use an analytical model of the ratio of network bandwidth and intra-node bandwidth to guide the selection of caching options. Figure~\ref{fig:network_compare} presents the measured bandwidth and latency of our testbed. Based on the characterization on the testbed, the dynamic caching needs to have at least 50\% cache hit rate to avoid performance loss.

\subsection{Memory Object Allocation}
The \tool interface for allocating a FAM-backed memory object supports two modes. First, anonymous mappings are supported by default to create empty pages. Second, if the application needs to pre-load saved data in a file into a FAM-backed memory object, the API also accepts a file name which will be opened on the server. Listing~\ref{lst:alloc} illustrates an example of allocating two memory objects in these two modes, respectively. %

\begin{lstlisting}[label=lst:alloc,basicstyle=\ttfamily\footnotesize,frame=single,language=C, 
framexleftmargin=0pt,caption={\tool user-level API examples.}]
void *anon_obj = SODA_alloc(&num_bytes, NULL);
void *file_obj = SODA_alloc(&num_bytes, file_name);
\end{lstlisting}

After a memory object is created, the application can use the returned pointer as a regular $malloc$-ed data. Under the hood, the \tool agent needs to reserve sufficient memory resources on the memory node side when a memory object is created. On the host side, the memory region belonging to a FAM-backed object is managed through Linux's \texttt{userfaultfd} (uffd) interface so that an accessed page is not in the host main memory, a notification will be forwarded to \tool host agent to trigger data transfer from the memory node. %

%% file: setup.tex
\section{Experimental Setup and Case Study}
\label{sec:setup}
For our experiments with \tool, we set the page size to 64~KB and the page buffer size to $1/3$ of the memory footprint. The dynamic DPU cache size is configured to store 1~GB data organized in an array of 1~MB chunks. The host is running Linux~4.18 and the DPU is running Linux~5.15. %

We perform a case study on common graph processing applications. Graph analytics is a powerful tool for modeling, analyzing, and optimizing complex systems in search engines, recommendation systems, and social media platforms. However, graph processing frameworks often need large memory capacity to be able to handle real-world graphs. This makes graph processing a promising target for disaggregated memory. We use Ligra~\cite{shun2013ligra}, a popular parallel graph processing framework, to utilize FAM by changing the graph construction routine to use the allocation APIs in \tool.

Ligra uses the sparse CSR format to enable efficient storage of large real-world graphs by splitting the vertex and edge data. In the original version, the full input data is read from disk into memory at initialization. After the modifications, the vertex and edge data structures are allocated and backed on a network-attached memory node. %
As the edge size is typically one order of magnitude larger than the vertex size (e.g., the datasets in this work range from 462~MB to 1.9~GB vertex data and 18~GB to 50~GB edge data), we use either static caching for vertex data or dynamic caching on the edge data in the DPU agent in the experiments.

We use five graph applications in Ligra for evaluation. \textit{Breadth-first Search (BFS)} constructs a search tree containing all nodes reachable from the initial source vertex on an input graph. \textit{PageRank (PR)} ranks each webpage based on the number and importance of inbound links. \textit{Radii} estimates the distance to the farthest vertex for each vertex in a graph. \textit{Betweenness centrality (BC)} finds the number of shortest paths passing through a vertex. \textit{Connected components (CC)} partitions an input graph into fully connected components. For input, we conduct our experiments on four real-world graphs, including \textit{com-friendster}, \textit{sk-2005}, \textit{twitter7}, and \textit{moliere\_2016}, from~\cite{davis2011university}. The characteristics of input graphs are summarized in Table~\ref{tab:input_graphs}. When configured to use static caching, the vertex data is chosen to be cached as it is relatively small and has high data accesses. %
We use 24~OpenMP threads to parallelize Ligra.

\begin{table}
    \centering
    \caption{A list of input graphs used for evaluation.}
    \label{tab:input_graphs}
    \begin{tabular}{llccc}
         \toprule
         Name & Type & $|V|$ & $|E|$ & $|E|/|V|$  \\
         \midrule
         friendster & social & 66~M & 3.6~B & 55 \\ %
         sk-2005 & web & 51~M & 1.9~B & 38 \\ %
         moliere & publications & 30~M & 6.7~B & 221 \\ %
         twitter7 & social & 42~M & 1.5~B & 35\\ %
         \bottomrule
    \end{tabular}

\end{table}

We split the nodes into compute nodes and memory nodes. For a compute node, we emulate the host as a thin compute node by using \texttt{cgroup} to limit the memory usage to 16~GB. For a memory node, it can use up to 256~GB DRAM to provide network-attached memory. On the DPU, the memory usage is limited to 1~GB. We configure the data transfer granularity between compute and memory nodes to be 64~KB. The host agent is pinned to NUMA node 2 based on the NUMA-aware optimization. %

To measure the network traffic volume, we utilized the network counters on the server.  We utilized network counters like \verb|port_xmit_data| in the mlx5 driver on the server to measure the network traffic. The difference between the counter value at the start and end of the experiment is measured as the number of transmitted 32-bit words.

%% file: results.tex
\section{Evaluation}
In this section, we evaluate the overall performance in single process graph processing applications and different caching and optimization strategies. We also evaluate multiple processes sharing a \tool service on SmartNIC.

\subsection{Overall Performance}

\begin{figure*}
    \centering
    \includegraphics[width=\linewidth]{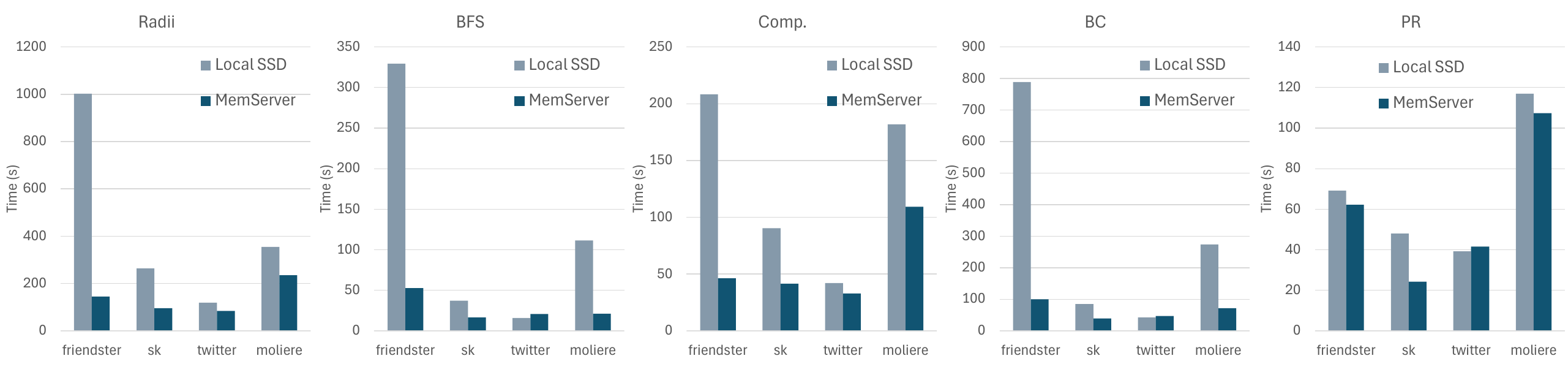}
    \caption{The performance comparison on five graph applications using node-local NVMe SSD and network-attached memory. Each graph application runs on four real-world graphs.}
    \label{fig:ssd}
\end{figure*}

Figure~\ref{fig:ssd} presents the performance of five applications running on 4 real-world graphs using a node-local NVMe SSD compared with a baseline memory server (without DPU). Among them, 17 cases achieved performance improvement using network-attached memory (the MemServer version) compared to the SSD. The remaining three cases are all from the twitter graph, where using node-local SSD is faster in BC, BFS and Radii applications by $10$--$20$\%. The generally high performance of remote memory compared to local SSD is consistent with previous works as RDMA network is faster than secondary storage~\cite{chen2023cowbird}. %
In addition to the performance gain and on-demand provisioning, network-attached memory nodes can be shared by compute nodes, which improves cluster-level performance per dollar over static provisioning~\cite{lim2009disaggregated,li2023pond}. %

\begin{figure*}
    \centering
    \includegraphics[width=\linewidth]{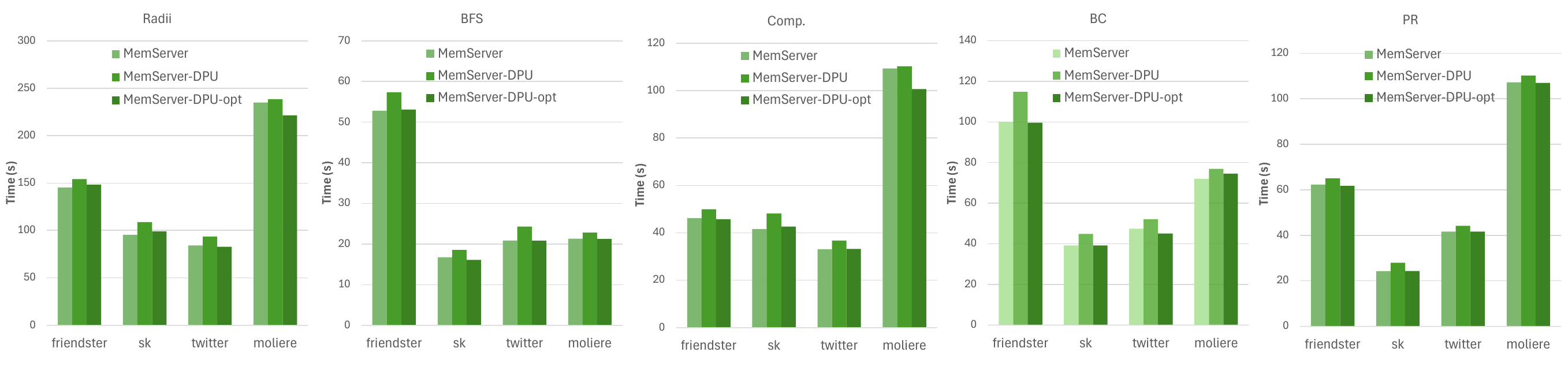}
    \caption{The overall performance of five graph applications on four real-world graphs. The three scenarios use baseline memory node or \tool DPU default and opt versions.}
    \label{fig:perf}
\end{figure*}

Next, we compare the overall performance of four network-attached memory versions in Figure~\ref{fig:perf}. The first version is the baseline memory server storing the data on the memory node, which is accessed directly from the host. The second and third versions use \tool for offloading tasks onto the DPU, where data caching is enabled in the third version (DPU opt). %

The DPU baseline is slower than the MemServer version by 1--14\% as presented in Figure~\ref{fig:perf}. Naively introducing the DPU into the data path without optimizations leads to increased latency per request and results in performance loss. However, the results are different with the DPU opt version. In six cases, the DPU-opt version is faster than the MemServer version by 1--9\%. Running Components and Radii applications on the Moliere brings the highest speedup compared to MemServer without offloading. Moliere is the largest dataset and also has the highest average vertex degree, indicating that \tool's performance is improved with larger datasets. In 10 cases, the DPU-opt and MemServer versions have similar performance. In the remaining four cases, DPU opt is slower than server-only by 2--4\%. In summary, offloading to SmartNIC naively brings little performance and our proposed optimizations are effective for a more performant implementation. %

Finally, the DPU-opt version achieves a speedup over the node-local SSD version in 18 out of 20 cases. The speedup ranges from 1.1 in Moliere-PageRank to 7.9 in the BC application on the friendster graph. Similar to the MemServer version, the node-local NVMe SSD version outperforms in two applications BFS and PageRank on the twitter graph by 10--20\%.

\subsection{Multi-process Graph Processing}
\begin{figure}[bt]
    \centering
    \includegraphics[width=\linewidth]{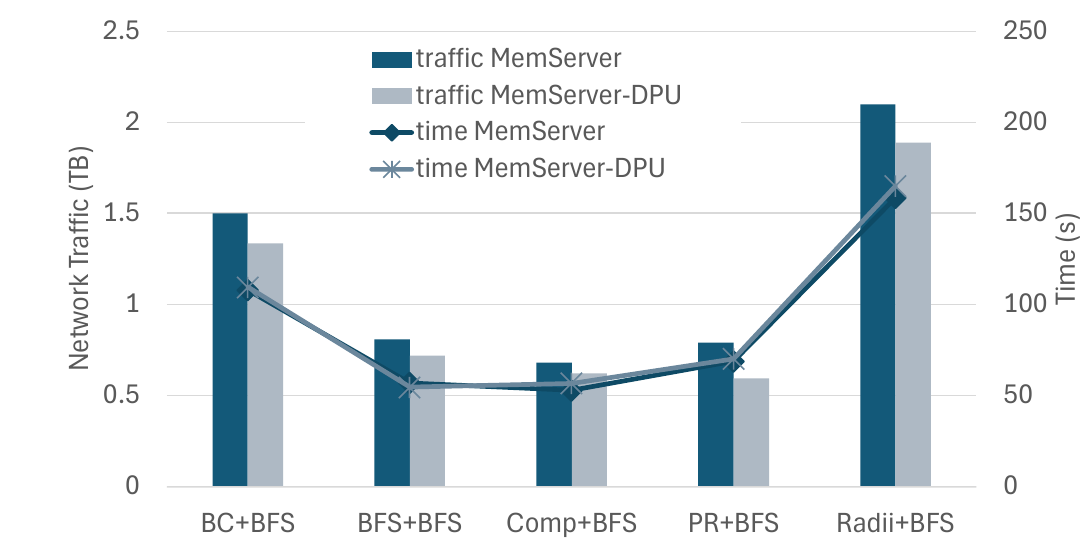}
    \caption{Network traffic reduction by offloading to \tool in five co-running graph applications on the com-friendster graph, as compared to the MemServer without offloading to DPU.}
    \label{fig:corun}
\end{figure}

We run multiple graph processes on the compute node concurrently, representing a realistic use case in data centers where multiple jobs share one node. In such use cases, several co-running processes may run different graph algorithms that analyze a large input graph to extract different insights.
With multiple processes on the same compute node, the SmartNIC and DPU agent are shared among multiple processes, so are the resources on the DPU. For instance, if they operate on the same dataset, the cache can be shared.

In this experiment, each application runs together with BFS as a background process on the com-friendster graph, using static caching. The execution time and network traffic compared with the server-only version are shown in Figure~\ref{fig:corun}. The network traffic is reduced by up to 25\% in PageRank and 9--11\% in the other applications. The results are similar to the traffic reduction in the single process case (not shown) and demonstrate the benefits of \tool scale to multi-process use cases. %

\subsection{Impact of Caching Options}
\begin{figure}[bt]
    \centering
    \includegraphics[width=\linewidth]{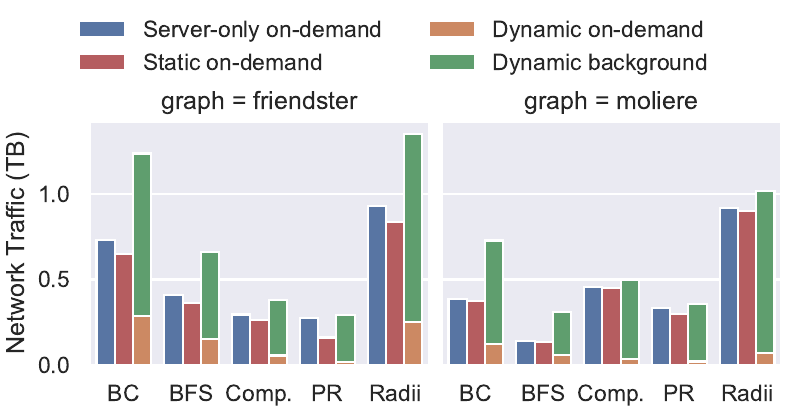}
    \caption{Network traffic in the server-only version and the DPU versions with two caching versions. Traffic is categorized into on-demand (critical path) or background (prefetching) traffic.}
    \label{fig:ondemand}
\end{figure}
\begin{figure}[bt]
    \centering
    \includegraphics[width=\linewidth]{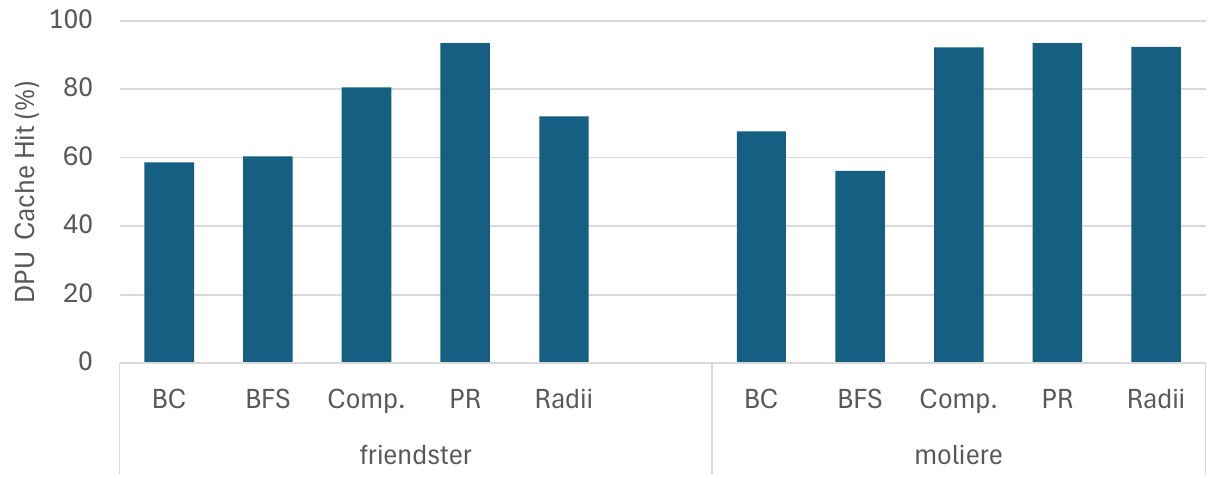}
    \caption{The cache hit rate of the dynamic caching strategy on DPU in five graph applications on two input graphs.}
    \label{fig:cache}
\end{figure}

We evaluate the two caching modes on com-friendster and Moliere. First, we consider the network traffic with caching compared to the server-only baseline, shown in Figure~\ref{fig:ondemand}. In com-friendster, static vertex caching reduces network traffic by 42\% in PageRank and 10--11\% in the other applications. In Moliere, the reduction is 10\% in PageRank and 2--3\% for the other applications. Static caching saves network traffic by reading the whole vertex data only once from the server into the DPU, which is amortized by later on-demand accesses which are managed locally by the DPU. The difference between the two graphs can be attributed to the difference in vertex degree -- Moliere has four times more edges per vertex and thus cached vertex accesses make up a smaller, although still critical, fraction of total network traffic.

On the other hand, dynamic edge caching may lead to increased network traffic. On the com-friendster graph, the increase ranges from 5\% in PageRank to 69\% in the BC application. In the Moliere graph, the increase ranges from 8\% in PageRank and 10\% in Components up to 117\% in BFS. Note however, that a significant fraction (about $76$-$93$\%) of the traffic in the dynamic caching mode has transformed from latency-critical on-demand transfers that run on the critical path, into prefetching transfers that run in the background. And, background traffic is less sensitive to network and server performance variability. %

We study the dynamic caching strategy by quantifying the cache hit rate (Figure~\ref{fig:cache}). On com-friendster, PageRank is the most predictable with 93\% hit rate, while BC is the least predictable with a hit rate of 61\%. On the Moliere graph, Components, PageRank and Radii have 92--93\% hit rate, while BC and BFS have only 56--68\% hit rate. The hit rate explains the difference in observed network traffic. With hit rates above 90\%, traffic increase is below 10\% and when the hit rate decreases to 56\%, the network traffic increases up to 117\% on Moliere-BFS. The results are consistent with the design in Section~\ref{sec:cache}, i.e., when the hit rate falls below a threshold, dynamic caching should be disabled on the DPU.

\subsection{Performance Breakdown}
\begin{figure}[bt]
    \centering
    \includegraphics[width=\linewidth]{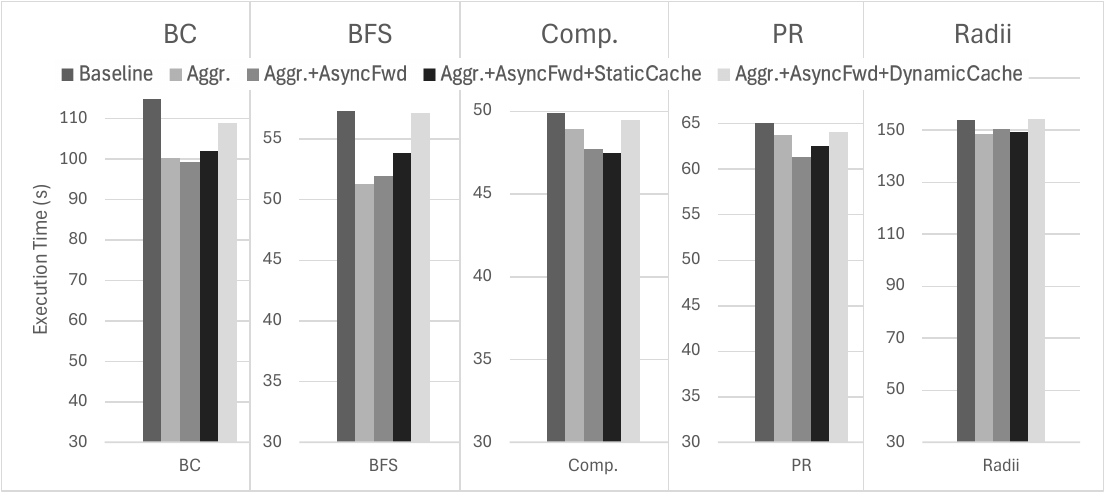}
    \caption{Performance breakdown of DPU optimizations in five graph applications on the com-friendster graph.}
    \label{fig:breakdown}
\end{figure}
We measure the performance of the baseline version with only specific optimizations enabled to study the effectiveness of optimizations. In the base version, all requests from the host are sent via the DPU, which forwards them to the server. The results on com-friendster are shown in Figure~\ref{fig:breakdown}.

Aggregation enables multiple requests to be processed in parallel, reducing the overhead of request handling. This provides a speedup over the base proxy from 15\% in BC to 2\% in Components. Asynchronous request forwarding decouples request receiving from forwarding by utilizing multiple DPU threads. Pipelining may improve throughput under high loads. In this experiment, the performance ranges from 4--3\% faster in PageRank and Components respectively, and the other applications within 1\% of the baseline. Caching does not improve running time. For static caching, the performance varies from unaffected in Components and Radii to 4\% slower in BFS. For dynamic caching, the performance varies from 3\% slower in Radii to 10\% slower in BFS. The dynamic cache introduces overhead in request handling due to lookups in and management of the cache table. 

In conclusion, aggregation and asynchronous forwarding provide modest performance improvement and should always be enabled. As for the two caching options, the performance in this experiment does not improve. However, caching can have other benefits such as reducing network traffic or mitigating network variability.

%% file: related.tex
\section{Related Works}
\textbf{Fabric-Attached Memory.} Works on RDMA-based disaggregated memory can be categorized into either OS-based~\cite{lim2009disaggregated,gu2017efficient,li2023pond} or application-specific, such as Seriema~\cite{mendes2022seriema}, an RDMA-based remote invocation framework for distributed data structures in C++1x focused on Monte-Carlo tree search, or FAM-Graph~\cite{zahka2022fam}, a DSL for graph processing applications to distribute data among local memory and network-attached memory which leverages specific application properties for optimizations. 

\textbf{SmartNIC Characterization.} Wei et al.~\cite{wei2023characterizing} present in-depth hardware details of the BlueField-2 and focus on characterizing different communication paths between the host, DPU, and the network. Liu et al.~\cite{liu2021performance} present a characterization study focused on the computational power of the BlueField-2 DPU, and also benchmark RDMA and DMA communication in the separated host mode.
Thostrup et al.~\cite{thostrup2022dbms} are focused on the use case of database management operations. Using a benchmark suite they find that many tasks are slower on a BlueField-2 DPU than on the host.

\textbf{SmartNIC for scientific applications.} Karamati et al.~\cite{karamati2022smarter} offload computations in the miniMD molecular dynamics code to BlueField-2. They restructure the algorithm to enable offloading of asynchronous background tasks to the DPU. BluesMPI~\cite{bayatpour2021bluesmpi} offloads the expensive non-blocking Alltoall collective operation in MPI to the BlueField-2 DPU to fully overlap communication and computation time. Ulmer et al.~\cite{ulmer2023extending} offload a particle-sifting data service onto the DPU for sorting and reorganizing particle data from a running simulation. Usman et al. developed ODOS~\cite{usman2023dpu} to enable offloading application tasks to a BlueField DPU using the OpenMP offloading programming model to improve productivity over low-level device programming or MPI schemes.

\textbf{SmartNIC for KVS.} SKV~\cite{sun2022skv} offloads the data replication in a distributed key-value store to a BlueField DPU to reduce the load on the host processor during write-heavy workloads. The evaluation shows that SKV can improve throughput by 14\% and reduce latency by 21\% compared to the baseline. 
Zhang et al.~\cite{zhang2023dow} design a DPU-offloaded KVS on disaggregated persistent memory by offloading small random writes to the DPU's DRAM.
iPipe~\cite{liu2019offloading} is a portable framework for programming both on-path and off-path SmartNICs using the actor model. The framework targets distributed applications such as key-value stores, transaction processing systems, and real-time analytics. Cowbird~\cite{chen2023cowbird} is an architecture for offloading remote memory accesses with use cases in key-value storage, implemented in programmable network switches or harvested spot VMs. In a KV application, Cowbird achieves the same throughput as fully in-memory execution.

%% file: conclusion.tex
\section{Conclusion}

Disaggregated memory is a promising solution for enhancing resource utilization in computing clusters.
It relies on software solutions to effectively manage various tasks to enable compute node memory expansion using network-attached memory.
In this work, we investigate the use of off-path SmartNICs to offload these tasks, such as monitoring, caching, and pipelining data movement between compute nodes and memory nodes. We propose a general SmartNIC-offloaded design incorporating NUMA awareness and caching optimizations tailored by system characteristics. Our prototype implementation, called \tool, is built on Nvidia's BlueField DPU. We evaluated \tool using five common graph applications on real-world graphs, employing single and multiple processes on compute nodes. \tool achieves up to a 7.9x speedup compared to node-local SSDs and reduces network traffic by up to 42\% while maintaining similar or better performance compared to a no-offloading baseline.

%% file: main-IEEE.bbl
\begin{thebibliography}{10}
\providecommand{\url}[1]{#1}
\csname url@samestyle\endcsname
\providecommand{\newblock}{\relax}
\providecommand{\bibinfo}[2]{#2}
\providecommand{\BIBentrySTDinterwordspacing}{\spaceskip=0pt\relax}
\providecommand{\BIBentryALTinterwordstretchfactor}{4}
\providecommand{\BIBentryALTinterwordspacing}{\spaceskip=\fontdimen2\font plus
\BIBentryALTinterwordstretchfactor\fontdimen3\font minus \fontdimen4\font\relax}
\providecommand{\BIBforeignlanguage}[2]{{%
\expandafter\ifx\csname l@#1\endcsname\relax
\typeout{** WARNING: IEEEtran.bst: No hyphenation pattern has been}%
\typeout{** loaded for the language `#1'. Using the pattern for}%
\typeout{** the default language instead.}%
\else
\language=\csname l@#1\endcsname
\fi
#2}}
\providecommand{\BIBdecl}{\relax}
\BIBdecl

\bibitem{shun2013ligra}
J.~Shun and G.~E. Blelloch, ``Ligra: a lightweight graph processing framework for shared memory,'' \emph{{ACM} {SIGPLAN} Notices}, vol.~48, 2013.

\bibitem{zhang2018graphit}
Y.~Zhang, M.~Yang, R.~Baghdadi, S.~Kamil, J.~Shun, and S.~Amarasinghe, ``Graphit: A high-performance graph dsl,'' \emph{Proceedings of the ACM on Programming Languages}, vol.~2, no. OOPSLA, 2018.

\bibitem{zivanovic2017main}
D.~Zivanovic, M.~Pavlovic, M.~Radulovic, H.~Shin, J.~Son, S.~A. Mckee, P.~M. Carpenter, P.~Radojkovi{\'c}, and E.~Ayguad{\'e}, ``Main memory in hpc: Do we need more or could we live with less?'' \emph{ACM Transactions on Architecture and Code Optimization (TACO)}, vol.~14, no.~1, 2017.

\bibitem{panwar2019quantifying}
G.~Panwar, D.~Zhang, Y.~Pang, M.~Dahshan, N.~DeBardeleben, B.~Ravindran, and X.~Jian, ``Quantifying memory underutilization in hpc systems and using it to improve performance via architecture support,'' in \emph{Proceedings of the 52nd Annual IEEE/ACM International Symposium on Microarchitecture}, 2019.

\bibitem{peng2020memory}
I.~Peng, R.~Pearce, and M.~Gokhale, ``On the memory underutilization: Exploring disaggregated memory on {HPC} systems,'' in \emph{2020 IEEE 32nd International Symposium on Computer Architecture and High Performance Computing (SBAC-PAD)}.\hskip 1em plus 0.5em minus 0.4em\relax IEEE, 2020.

\bibitem{michelogiannakis2022case}
G.~Michelogiannakis, B.~Klenk, B.~Cook, M.~Y. Teh, M.~Glick, L.~Dennison, K.~Bergman, and J.~Shalf, ``A case for intra-rack resource disaggregation in hpc,'' \emph{ACM Transactions on Architecture and Code Optimization (TACO)}, vol.~19, no.~2, pp. 1--26, 2022.

\bibitem{li2023analyzing}
J.~Li, G.~Michelogiannakis, B.~Cook, D.~Cooray, and Y.~Chen, ``Analyzing resource utilization in an hpc system: A case study of nersc’s perlmutter,'' in \emph{International Conference on High Performance Computing}.\hskip 1em plus 0.5em minus 0.4em\relax Springer, 2023, pp. 297--316.

\bibitem{vazhkudai2018design}
S.~S. Vazhkudai, B.~R. De~Supinski, A.~S. Bland, A.~Geist, J.~Sexton, J.~Kahle, C.~J. Zimmer, S.~Atchley, S.~Oral, D.~E. Maxwell \emph{et~al.}, ``The design, deployment, and evaluation of the coral pre-exascale systems,'' in \emph{SC18: International Conference for High Performance Computing, Networking, Storage and Analysis}.\hskip 1em plus 0.5em minus 0.4em\relax IEEE, 2018, pp. 661--672.

\bibitem{lim2009disaggregated}
K.~Lim, J.~Chang, T.~Mudge, P.~Ranganathan, S.~K. Reinhardt, and T.~F. Wenisch, ``Disaggregated memory for expansion and sharing in blade servers,'' \emph{ACM SIGARCH computer architecture news}, 2009.

\bibitem{gu2017efficient}
J.~Gu, Y.~Lee, Y.~Zhang, M.~Chowdhury, and K.~G. Shin, ``Efficient memory disaggregation with infiniswap,'' in \emph{14th USENIX Symposium on Networked Systems Design and Implementation (NSDI 17)}, 2017.

\bibitem{li2023pond}
H.~Li, D.~S. Berger, L.~Hsu \emph{et~al.}, ``Pond: Cxl-based memory pooling systems for cloud platforms,'' in \emph{Proceedings of the 28th ACM International Conference on Architectural Support for Programming Languages and Operating Systems}, 2023.

\bibitem{wahlgren2023quantitative}
J.~Wahlgren, G.~Schieffer, M.~Gokhale, and I.~Peng, ``A quantitative approach for adopting disaggregated memory in hpc systems,'' in \emph{Proceedings of the International Conference for High Performance Computing, Networking, Storage and Analysis}, 2023, pp. 1--14.

\bibitem{zahka2022fam}
D.~Zahka and A.~Gavrilovska, ``Fam-graph: Graph analytics on disaggregated memory,'' in \emph{2022 IEEE International Parallel and Distributed Processing Symposium (IPDPS)}.\hskip 1em plus 0.5em minus 0.4em\relax IEEE, 2022.

\bibitem{wei2023characterizing}
X.~Wei, R.~Cheng, Y.~Yang, R.~Chen, and H.~Chen, ``Characterizing off-path {SmartNIC} for accelerating distributed systems,'' in \emph{17th USENIX Symposium on Operating Systems Design and Implementation (OSDI 23)}, 2023.

\bibitem{wahlgren2022evaluating}
J.~Wahlgren, M.~Gokhale, and I.~B. Peng, ``Evaluating emerging cxl-enabled memory pooling for hpc systems,'' in \emph{2022 IEEE/ACM Workshop on Memory Centric High Performance Computing (MCHPC)}.\hskip 1em plus 0.5em minus 0.4em\relax IEEE, 2022, pp. 11--20.

\bibitem{de2022noise}
D.~De~Sensi, T.~De~Matteis, K.~Taranov, S.~Di~Girolamo, T.~Rahn, and T.~Hoefler, ``Noise in the clouds: Influence of network performance variability on application scalability,'' \emph{Proceedings of the ACM on Measurement and Analysis of Computing Systems}, vol.~6, no.~3, 2022.

\bibitem{liu2021performance}
J.~Liu, C.~Maltzahn, C.~Ulmer, and M.~L. Curry, ``Performance characteristics of the {BlueField-2 SmartNIC},'' \emph{arXiv preprint arXiv:2105.06619}, 2021.

\bibitem{karamati2022smarter}
S.~Karamati, C.~Hughes, K.~S. Hemmert, R.~E. Grant, W.~W. Schonbein, S.~Levy, T.~M. Conte, J.~Young, and R.~W. Vuduc, ``{“Smarter”} {NICs} for faster molecular dynamics: a case study,'' in \emph{2022 IEEE International Parallel and Distributed Processing Symposium (IPDPS)}.\hskip 1em plus 0.5em minus 0.4em\relax IEEE, 2022.

\bibitem{ewais2023disaggregated}
M.~Ewais and P.~Chow, ``Disaggregated memory in the datacenter: A survey,'' \emph{IEEE Access}, vol.~11, pp. 20\,688--20\,712, 2023.

\bibitem{kalia2016design}
A.~Kalia, M.~Kaminsky, and D.~G. Andersen, ``Design guidelines for high performance {RDMA} systems,'' in \emph{USENIX Annual Technical Conference}, 2016.

\bibitem{davis2011university}
T.~A. Davis and Y.~Hu, ``The university of florida sparse matrix collection,'' \emph{ACM Transactions on Mathematical Software (TOMS)}, vol.~38, no.~1, 2011.

\bibitem{chen2023cowbird}
X.~Chen, L.~Yu, V.~Liu, and Q.~Zhang, ``Cowbird: Freeing cpus to compute by offloading the disaggregation of memory,'' in \emph{Proceedings of the ACM SIGCOMM 2023 Conference}, 2023.

\bibitem{mendes2022seriema}
H.~Mendes, B.~Wiedenbeck, and A.~O'Neill, ``Seriema: Rdma-based remote invocation with a case-study on monte-carlo tree search,'' in \emph{2022 IEEE 34th International Symposium on Computer Architecture and High Performance Computing (SBAC-PAD)}.\hskip 1em plus 0.5em minus 0.4em\relax IEEE, 2022, pp. 11--20.

\bibitem{thostrup2022dbms}
L.~Thostrup, D.~Failing, T.~Ziegler, and C.~Binnig, ``A {DBMS}-centric evaluation of {BlueField} {DPUs} on fast networks,'' in \emph{13th International Workshop on Accelerating Analytics and Data Management Systems Using Modern Processor and Storage Architectures}, 2022.

\bibitem{bayatpour2021bluesmpi}
M.~Bayatpour, N.~Sarkauskas, H.~Subramoni, J.~Maqbool~Hashmi, and D.~K. Panda, ``{BluesMPI}: Efficient {MPI} non-blocking {Alltoall} offloading designs on modern {BlueField} smart {NICs},'' in \emph{International Conference on High Performance Computing}.\hskip 1em plus 0.5em minus 0.4em\relax Springer, 2021.

\bibitem{ulmer2023extending}
C.~Ulmer, J.~Liu, C.~Maltzahn, and M.~L. Curry, ``Extending composable data services into {SmartNICs},'' in \emph{2023 IEEE International Parallel and Distributed Processing Symposium Workshops (IPDPSW)}.\hskip 1em plus 0.5em minus 0.4em\relax IEEE, 2023.

\bibitem{usman2023dpu}
M.~Usman, S.~Iserte, R.~Ferrer, and A.~J. Pe{\~n}a, ``{DPU} offloading programming with the openmp api,'' in \emph{Proceedings of the SC'23 Workshops of The International Conference on High Performance Computing, Network, Storage, and Analysis}, 2023, pp. 884--891.

\bibitem{sun2022skv}
S.~Sun, R.~Zhang, M.~Yan, and J.~Wu, ``{SKV}: A {SmartNIC}-offloaded distributed key-value store,'' in \emph{2022 IEEE International Conference on Cluster Computing (CLUSTER)}.\hskip 1em plus 0.5em minus 0.4em\relax IEEE, 2022.

\bibitem{zhang2023dow}
Y.~Zhang, G.~Li, J.~Wan, J.~Wang, J.~Li, T.~Yao, H.~Wu, and D.~Wang, ``Dow-kv: A dpu-offloaded and write-optimized key-value store on disaggregated persistent memory,'' in \emph{2023 IEEE International Conference on Cluster Computing (CLUSTER)}.\hskip 1em plus 0.5em minus 0.4em\relax IEEE, 2023.

\bibitem{liu2019offloading}
M.~Liu, T.~Cui, H.~Schuh, A.~Krishnamurthy, S.~Peter, and K.~Gupta, ``Offloading distributed applications onto {smartNICs} using {iPipe},'' in \emph{Proceedings of the ACM SIGCOMM '19}.\hskip 1em plus 0.5em minus 0.4em\relax ACM, 2019.

\end{thebibliography}
